# Chemputer and Chemputation - A Universal Chemical Compound Synthesis Machine

Leroy Cronin*, Sebastian Pagel, and Abhishek Sharma

*School of Chemistry, Advanced Research Centre, University of Glasgow, Glasgow, G11 6EW UK, www.croninlab.com email: lee.cronin@glasgow.ac.uk

**Abstract**

Chemputation reframes synthesis as the programmable execution of reaction code on a universally re-configurable hardware graph. Here we prove that a chemputer equipped with a finite—but extensible—set of reagents, catalysts and process conditions, together with a "chempiler" that maps reaction graphs onto hardware, is universal: it can generate any stable, isolable molecule in finite time and in analytically detectable quantity, provided real-time error correction keeps the per-step fidelity above the threshold set by the molecule's assembly index. The proof is constructed by casting the platform as a Chemical Synthesis Turing Machine (CSTM). The CSTM formalism supplies (i) an eight-tuple state definition that unifies reagents, process variables (including catalysts) and tape operations; (ii) the Universal Chemputation Principle; and (iii) a dynamic-error-correction routine ensuring fault tolerant execution. Linking this framework to assembly theory strengthens the definition of a molecule by demanding practical synthesizability and error correction becomes a prerequisite for universality. We validate the abstraction against >100 χDL programs executed on a modular chemputer rigs spanning single step to multi-step routes. Mapping each procedure onto CSTM shows that the cumulative number of unit operations grows linearly with synthetic depth. Together, these results elevate chemical synthesis to the status of a general computation: algorithms written in χDL are compiled to hardware, executed with closed-loop correction, and produce verifiable molecular outputs. By formalising chemistry in this way, the chemputer offers a path to shareable, executable chemical code, interoperable hardware ecosystems, and ultimately a searchable, provable atlas of chemical space.

**Significance Statement**

Chemical synthesis is still performed today much like bespoke craftsmanship—each target molecule demands specialized equipment, ad hoc protocols, and labor intensive trial and error. We demonstrate that this is not a fundamental limitation of chemistry by formalising a Chemical Synthesis Turing Machine (CSTM). With a practical modular chemputer we prove that a single, re configurable hardware graph equipped with real time error correction can, in principle, construct any stable, isolable molecule in analytically detectable quantities. We explore this universality with assembly theory, establishing the first quantitative bridge between a molecule's intrinsic information content (assembly index) and the fault tolerant resources required for its synthesis.

## Introduction

Turing completeness is a concept from theoretical computer science that defines the ability of a computational system to perform any computation that can be done by a Turing machine [1–3]. For a system to be Turing-complete, it must have the capability to simulate a Turing machine. This means it can execute any algorithm, given sufficient time and memory, and solve any problem that is computationally solvable. Turing completeness is a foundational concept in understanding the limits of what can be computed. In essence, if a programming language or computational system is Turing complete, it can, in theory, perform any computation that a computer can, assuming no constraints on resources like time and memory. Expanding this concept to the realm of chemistry involves envisioning chemical systems that can perform operations that are in a way analogous to a Turing machine[4]. Here we explore this idea where chemical reactions are used to undergo programmable transformations in a device we call a chemputer[5–9]. The chemputer is designed to automate and control chemical reactions with high precision[10]. It uses a combination of hardware and software to carry out complex sequences of chemical processes[11]. By programming these sequences, the chemputer can perform tasks that require conditional logic, loops, and the manipulation of data—key components of Turing completeness.

Practically speaking, there are now many notable examples of chemistry automation with a wide range of chemical reactions, and therefore, represent hardware-specific chemical processes. For instance, the synthesis of sequence-defined biopolymers whose syntheses already follow deterministic, stepwise logic with some feedback control. A great example is modern solid-phase peptide synthesis,[12,13] which is routinely used to construct complex peptides using protected amino-acid cartridges and inline deprotection checks, while on-solid-phase cleavage is triggered only once conversion has occurred. The same approach governs automated oligonucleotide production[14]. Perhaps the most ubiquitous reaction to be encoded for small molecules has been the amide-bond formation,[15] and this is now followed by Suzuki–Miyaura[16], Buchwald–Hartwig,[17] and Sonogashira[18] reactions. Other examples include microfluidic systems for droplet-based chemistry[19], DNA encoded libraries[20] and other combinatorial chemistry approaches[21]. In addition, recently there has been an explosion of work using flow systems for chemical synthesis[18,22] as well as a vast number of digital chemistry or self-driving lab endeavours[23,24,25,26] many of which are aiming for digital synthesis[27–29,30], real-time monitoring and optimisation,[31–34] mobility[35,36], autonomy[37,38]. There are clear advantages for investment in such systems for the control of highly exothermic reactions, process optimisation, and exploring new inline sensors. However, in all these examples, they are all encoded from the hardware up, meaning that a fully abstract approach to programming the chemistry has not been possible. These challenges are important because whilst there are use cases for all these approaches, the real challenge is to find a universal route to in principle encode all of chemistry at an abstract level so that it can then be run on chemically agnostic devices. Only then will the promise of digital chemistry become a reality.

Here we present a universal approach to programming and executing chemical operations that will lay the foundations for Chemputation, unifying the programming of chemistry across all of chemical synthesis, design, discovery and automation. By having a universal, 'Turing complete' standard, it will be possible to embrace the many different approaches into a single programmable paradigm. This will allow aspects of provability, interoperability, defining an entirely new ecosystem. With a universal system, teams will be able to develop a common programming standard, develop interoperable hardware modules, produce systems that are able to reproduce each other's results, build a repository of both negative and positive reaction data, and, finally, publish executable chemical code. We defined a chemputer as a system that can execute a standard chemical code to make a wide range of different molecules and the process of running the machine with the code to get the chemical outputs as chemputation.

**1. Foundation of the Chemputer**

Since the late-1960s progress in modular robotics, low-cost multimodal sensing, and data-driven route design has propelled the field of chemical synthesis from task-specific robots—peptide assemblers, DNA synthesisers, high-throughput flow loops—toward fully programmable chemical platforms. Yet, as described above most current systems remain constrained to a narrow reaction manifold and typically lack the capacity to rewire their own hardware topology in response to a new synthesis plan. To appreciate the qualitative leap from single-use automation to a universal chemical programming language,[39] it helps to recall the gulf between an abacus and a programmable computer, see Figure 1. The abacus performs fixed arithmetic by sliding beads along rails—it is fast and reliable, but it cannot be coaxed into factoring integers, searching a database, or rendering graphics without physically altering its structure. Its function set is frozen in hardware. By contrast, a computer's power stems from a universal instruction set. Here, the processes of load, add, branch, and store are integral. These few op-codes, scripted in software, can realise any computable routine, limited only by time and memory. In contrast to the single use chemical synthesis machines, c.f. the abacus, where the reaction process is mostly hard coded, the chemputer concept generalises these efforts as it treats synthetic chemistry itself as a form of computation.

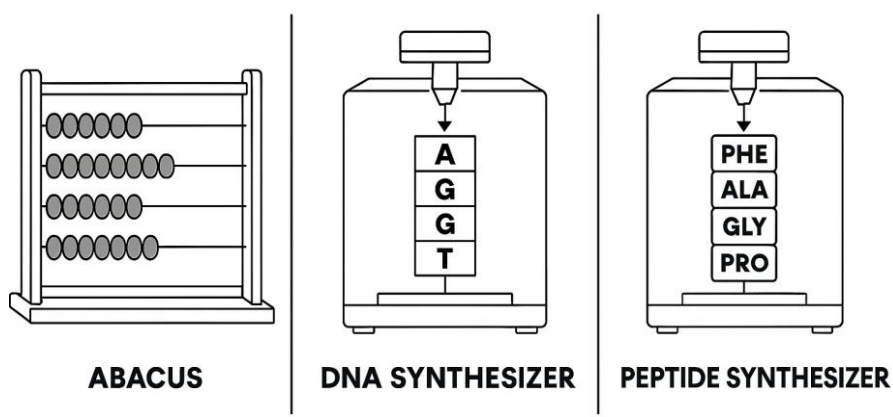

**Figure 1.** The abacus, DNA and peptide synthesizer are both examples of limited programmable machines. The abacus for numerical calculations and the DNA and peptide synthesizer for making sequences of DNA and peptides respectively.

This is because the reagents are data structures, reaction conditions are control flow, and hardware modules act as the physical instantiation of op-codes. By introducing a chempiler—a compiler that converts abstract reaction graphs into executable hardware graphs—the chemputer aims not merely to accelerate known protocols but to search, optimise, and execute entirely new routes on demand, in the same way a universal Turing machine can execute any algorithm expressible in its instruction set.

In this paper we formalise chemputation by (i) defining an abstract extended chemical state machine that captures reagents, catalysts, and process conditions as state variables and show how resources on a graph can be used to instantiate a wide range of chemical processes; (ii) proving a Universal Synthesis Theorem that any stable, isolable molecule can be reached through a finite sequence of such state transitions; and (iii) embedding dynamic error-correction routines that guarantee robustness in the face of real-world deviations. This foundation lays the groundwork for a future in which chemical manufacture is as programmable and portable as software is today, and it also provides a connection between assembly theory[40] and chemputation. This is because the assembly index is a measure of molecular complexity,[41] or the minimum number of constraints required to construct the molecule by considering bonds as building blocks. This work also demonstrates how the concept of the assembly index and the copy number play a profound role in understanding what is synthetically accessible and can be detected using analytical chemistry techniques.[42,43]

## 2. The Concept

The concept of a chemputer as a universal chemical synthesis machine posits that it can instantiate any feasible chemical synthesis, see Figure 2. We outline the proof for the universality of the chemputer, demonstrating that it can synthesize any target compound within the chemical space defined by the provided parameters. To prove the universality of the chemputer, we need to demonstrate that it can conduct any feasible chemical synthesis at the most abstract level. This involves showing that the transformation function δ can be used to map all chemical reactions possible under the defined reagents, process conditions, and catalysts (it has been suggested that catalysts might themselves be viewed as a type of constructor)[44,45]. Furthermore, we incorporate the mechanisms of dynamic error correction[46,47] during synthesis[22,48] and the use of universally configurable hardware to support complex chemical processes through a chempiling function. This allows the abstraction to be implemented at the module and device level for chemical synthesis.

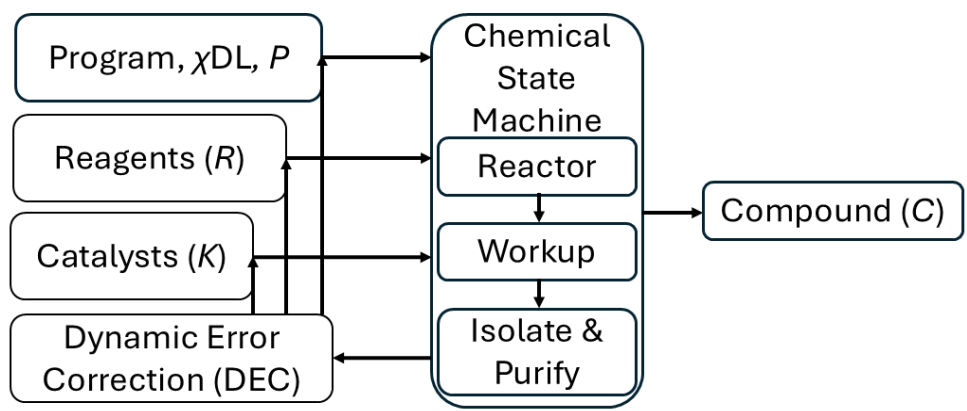

**Figure 2:** A schematic of one possible Chemical Synthesis Turing Machine (CSTM). The inputs are the Reagents ($R$), and the chemical program or χDL file[8] contains details of the process conditions ($P$) which include any catalysts ($K$), and code to run the hardware. The output is the pure target compounds ($C$) from the chemical state machine which includes a reactor, workup, isolate and purify system. Dynamic error correction (DEC) can adjust the process variables, reagents, and catalysts to ensure the target compound is produced reducing the error rate per step as much as possible maximising the yield of the compound.

## 3. The Turing Machine Abstraction of the Chemputer

To build the abstraction of the Chemputer, we need to set up the abstraction of the Chemical Synthesis Turing Machine (CSTM). This comprises an infinite tape where each space on the tape is a vessel, and each vessel can be one of three types of being either empty, filled, or active. The vessels can be subjected to the four primitives of Adding Matter [AM], Subtracting Matter [SM], Adding Energy [AE], or Subtracting Energy [SE] using the head. From these four primitives, any unit operation can be emulated over the entirety of chemical synthesis, see Table 1.

| Unit operations | Primitive Sequence | Notes |
| --- | --- | --- |
| Liquid-liquid extraction | AM → AE → SM | Add, mix, separate |
| Drying | AE→SM | Heat, remove water |
| Crystallisation | AE→SE→SM | Heat, cool, filter |

| Distillation | AE→ SM → SE → AM | Heat, remove vaporize, cool, collect |
|---|---|---|
| Hot reaction; cold reaction | AM→AE; AM→SE | Add matter then heat or cool |
| Sublimation | SM→AE→SE→AM | Add vacuum, heat, cool window, collect from window |

**Table 1.** Example unit operations from chemistry expressed in the primitives.

To fully express the machinery for chemical synthesis, we define the Chemical Synthesis Turing Machine, CSTM, as the 8-tuple:

$$C = \langle Q, \Sigma_R, \Sigma_P, \Gamma, b, \delta, q_0, H \rangle \text{ (eq 1.) where}$$

$Q$ is a finite set of states and $q_o \in Q$ is the initial state;

$\Sigma_R$ is a finite reagent alphabet;

$\Sigma_P$ is a process alphabet;

$\Gamma = (\Sigma_R \times \Sigma_P) \cup \{b\}$ is the tape alphabet whose blank symbol is $b$;

$\delta: Q \times \Gamma \rightarrow Q \times \Gamma \times \{Left, Right, N\}$ where the transition function, realised physically by the primitives AM, SM, AE, SE;

$H = \{q_{out}, q_{uout}, q_{nout}, q_{fail}\} \subseteq Q$ is halting set with,

- $q_{out}$ is a successful termination identical to a previously characterised chemputation;
- $q_{uout}$ halting after a theoretically predicted but as yet unoptimized outcome;
- $q_{nout}$ marking the discovery of a genuinely novel transformation;
- $q_{fail}$ indicating a chemically unrecoverable termination.

In every halting case the final tape encodes a complete laboratory trace. For $q_{uout}$ and $q_{nout}$ an external optimization module (DEC) may launch a non-deterministic exploration of $(\Sigma_R \times \Sigma_P)$ to improve yield or generate new reaction rules before the exact trace is committed to the rule database, see Figure 3. When a either a $q_{uout}$ or $q_{nout}$ is repeated it becomes $q_{out}$.

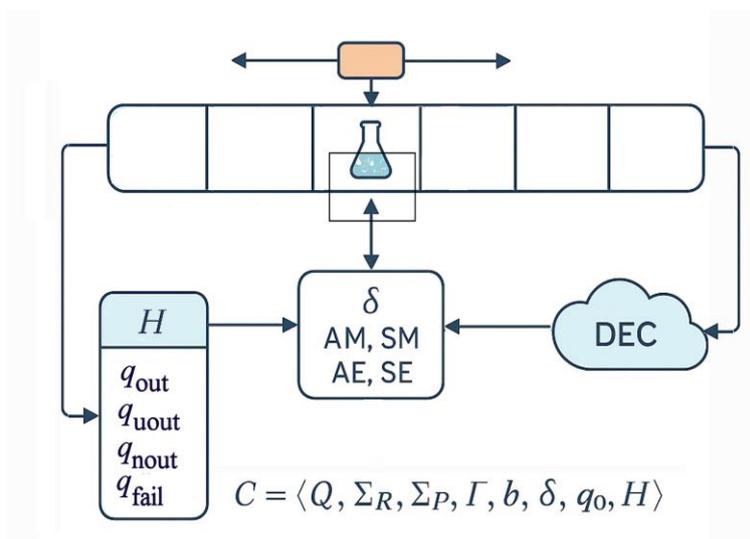

**Figure 3.** A schematic depicting the abstraction of the Universal Chemical Synthesis Turing Machine (CSTM). This machine is controlled by a finite state controller and the head addresses the cells on the tape by either adding matter, subtracting matter, adding energy, or subtracting energy. The system halts when one of four conditions is satisfied. Dynamic error correction (DEC) can be used to improve the outcomes.

The implementation of the CSTM is designed such that it preserves mass balance see Supplementary Data. i.e. the result of the transformations does not lose matter and hence stoichiometry. When matter is subtracted, the contents are removed to another vessel and if waste, to the waste vessel. Overall, for the synthesis of a compound from the set of all possible compounds, we can show that $\forall\ c \in C$, $\exists$ CSTM program $X_c$ such that out($X_c$) = c $\wedge$ $X_c$ halts. So, it is now straightforward to construct an archetypal CSTM schema that can be used to react reagents A and B together. By setting the tape of the machine to be equivalent to a series of chemical vessels, it is possible to build the machine that can use the head to address the locations on the tape to conduct one of the four-unit operations, see supplementary information. Here the cells can be considered to be infinite as long as material can be removed from the cells and the cells instantiated ready for further operations, see supplementary information and SV1.

## 4. Definitions

**Chemputer** and **Chemputation**: A system that runs the code to do the chemistry is the chemputer and the process of running the code is chemputation.

**Reagent Space (*R*)**: A finite working set drawn from the space of all possible chemical reagents, including all chemical elements and basic compounds.

**Process Conditions (*P*)**: A set of environmental parameters (e.g., temperature, pressure, solvent/gas conditions, energy input type) that influence the outcome of reactions. This includes **Catalysts (*K*)**: A set of substances that alter the reaction pathways or rates without being consumed in the process.

**Target Compounds (*C*)**: The set of desired products or output compounds, typically molecules. Such molecules can be defined as an electrically neutral entity consisting of more than one atom. In addition, the molecule must correspond to a depression on the potential energy surface that is deep enough to confine at least one vibrational state. Finally, the molecule should be accessible by chemputation, with a sufficiently large amount of the molecule synthetically accessible to be detected.

**Universally Configurable Hardware (*H*)**: A hardware platform that can be dynamically reconfigured to execute various chemical synthesis processes. In the chemputer, the system is constrained by a finite number of reagent input vessels, reaction vessels, and product output vessels, $V_R$, $V_P$, $V_O$ respectively; however, the tape is conceptually infinite since the vessels can be instantiated serially by re-using vessels and off-loading intermediates. This means that any chemical synthesis is realizable if it can be completed within these finite resources. The configuration is represented as a graph $G = (V,E)$, where: $V$ is a set of nodes representing hardware components (e.g., reactors, mixers, sensors) and $E$ is a set of edges representing connections between components, defining the flow of matter, energy, and information, see Figure 4.

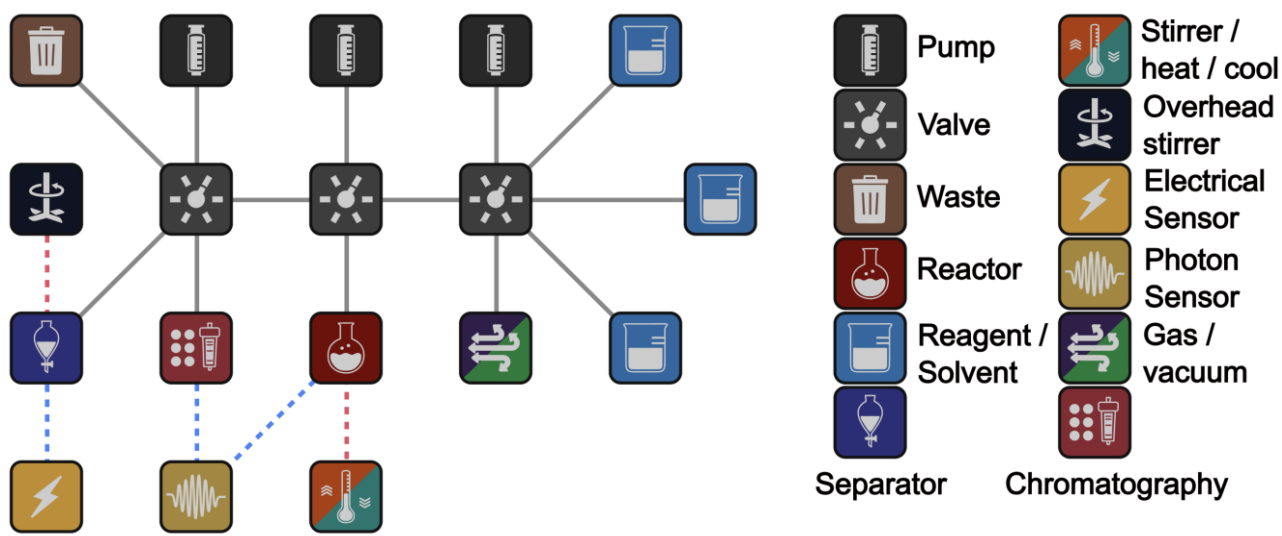

**Figure 4:** A general-purpose graph for the chemputer that can be practically implemented in the laboratory. The modules are shown on the right. The reagents, pumps, and valves are needed with the reactor to setup the reaction. The stirrer / heater / cooler is used to control the reaction. The reaction work-up uses a separator and a conductivity sensor. The combination of spectroscopic (photon sensor) on the reactor and chromatography output can allow real time feedback to help optimize both the reaction yield, purity, and selectivity.

This graph has the basic hardware resources for chemical reactions, work-up, isolation, and purification and the modules shown are practical implementations of the four Chemputation primitives: add matter, subtract matter, add energy, subtract energy. As such a system such as that shown in Fig 4. is capable of a wide range of chemputations. Of course there will always be practical limitations but the system shown above is not process or chemical reaction limited.

**Dynamic Error Correction (DEC)**: A mechanism embedded within each step of the synthesis process, enabling real-time detection and correction of errors, ensuring the accuracy of each transformation including the discovery of new transformations, before proceeding to the next step.

**Chempiling Function ($\chi$)**: The process of translating a synthesis pathway $\sigma$ into a corresponding hardware configuration $G(H)$ that can execute the synthesis process.

**5. Definition of a molecule, the assembly index, and the role of error correction.**
Chemical synthesis is inherently prone to error, and as molecular complexity increases, achieving error-free assembly becomes exponentially more difficult. To formalize this constraint, we expand the traditional definition of a molecule by incorporating practical synthetic accessibility.

Conventionally, a molecule is defined[49] as a finite set of nuclei and electrons occupying a stable local minimum on the Born-Oppenheimer potential energy surface. Here, we refine this by requiring that the molecule must also be realisable: it must be possible to produce enough perfect copies to detect the molecule experimentally, despite finite synthesis resources and inevitable errors.
Specifically, a molecule must satisfy the condition:

$$N \geq N_{min} = \frac{\varphi}{\prod_{k=1}^{a_i}(1-\varepsilon_k)} \qquad (eq.\ 2)$$

Where $N_{min}$ is the minimum number of copies that must be synthesised and $N$ is the number of perfect copies
$\varphi$ is the minimum number of perfect copies required for reliable detection (typically $10^6$-$10^8$),
$a_i$ is the assembly index[41,42] the minimal number of logical steps needed to construct the molecule,
$\varepsilon_k$ is the error probability at each assembly step.
$N_{perfect}$ is the number of molecules produced given the intrinsic error rate for each step and is equal to $\prod_{k=1}^{a_i}(1-\varepsilon_k)$.

For simplicity, assuming a constant error rate $\varepsilon$ across all steps, this expression reduces to:

$$N \geq N_{min} = \frac{\varphi}{(1-\varepsilon)^{a_i}} \qquad (eq.\ 3)$$

Thus, as the assembly index $a_i$ increases, or as the per-step fidelity declines, the number of molecules required to ensure detection grows exponentially. To provide intuition: each synthetic operation is like placing a brick when building a fragile structure. If each brick has a small chance of being misplaced, the probability of completing the structure without error declines rapidly as the number of bricks increases.

**6. Axioms and Lemmas**

For the CSTM to operate universally we need to introduce five axioms (A) and three lemmas (L): A1: Conservation of matter; A2: Finite reaction time; A3: Stability of elements found in *R* under standard conditions; A4: Every compound has a shortest path, defined by the assembly index, $a_i$, which is defined as the shortest path to assemble the compound from fundamental building blocks, allowing only binary combination of parts, and allowing reuse of parts; A5: Detectability constraint; $c \in C$, a synthesis is considered realisable only if the expected flawless-copy count satisfies $N \geq N_{min}$. L1: For any $c \in C$ there exists a finite sequence of transformations $\sigma$ and fine copy number such that *N* such that $\sigma$ executed *N* times satisfies A5. Proof: By the definition of *C* and finite reaction time axiom with a sequence of transformations $\sigma$; L2: For any $c \in C$ there exists a shortest path to construct c on a graph that only uses the building blocks found within *c*, allowing recursion. Proof: By the definition of *C*; L3: For any transformation function $t \in \delta$ can be decomposed into a finite sequence of elementary reactions. Proof: By the nature of chemical reactions and the conservation of matter.

**7. Assumptions and formalisation**

1. **Existence of a Universal Set-Up**: This demonstrates that the chemputer can implement any feasible chemical synthesis, showing that the function $\delta$ is sufficiently general to account for all chemical reactions possible under the reagents given, process conditions, and catalysts.
2. **Construction of Synthesis Pathway**: For each target compound *c*, a sequence $\sigma$ of transformations from initial reagents $R_0$ to *c* can be constructed. This construction must account for all intermediate transformations and ensure that $\sigma$ is valid under *P*, and *K*.
3. **Verification of Stability**: This verifies that for the resulting compound *c*, the stability condition $S(c)$ is satisfied.
4. **Dynamic Error Detection and Correction**: The chemputer can detect errors in real-time during each step of the synthesis by continuously monitoring the reaction progress and comparing the actual outcome with the expected result. Upon detecting an error during any synthesis step, the chemputer applies corrective steps immediately, either reverting to a previous state or adjusting the process to ensure the synthesis remains on track.

5. **Universality and Completeness**: This proves that for any $c \in C$, there exists a pathway $\sigma$ and a stable outcome, demonstrating the universality of the chemputer as a synthesis device, including error detection and correction at each synthesis step.

A molecule $c \in C$ is considered realizable in the chemputer is:

$$S(c) : \text{Stability } c \in C \text{ such that } c \text{ is isolable and stable} \quad (\text{eq. 4})$$

$\sigma$ exist such that $\sigma(R_0,...,R_n) = c$ and

$\sigma$ produces at least $N_{min}$ perfect copies as defined in eq. 2.

The stability condition $S(c)$ ensures that the resulting compound $c$ is stable and can be isolated, i.e., $S(c)$ must hold true for the synthesis to be considered successful. However, the synthesis may or may not utilize unstable reaction intermediates that could be isolated for some period of time. For the target to be produced there is a transformation function ($\delta$) done by the CSTM where

$$\delta \text{ maps } Q \times \Gamma \text{ to } Q \times \Gamma \times \{Left, Right, N\} \rightarrow C \quad (\text{eq. 5})$$

The transformation function $\delta$ from the CSTM defines the emergent property we conventionally call the reaction rule which is the resultant outcome when reagents $R$ are added under the process conditions $P$, in the presence of catalysts $K$ to give the output compounds $C$. The transformation function can be used to predict how the reagent graphs $R$ can be transformed into the product graphs $C$ as graph transformations between the reagents $R$. To get to the product we need to achieve the construction of synthesis pathways ($\sigma$) so for any target compound $c \in C$, we construct a pathway $\sigma$ such that:

$$\sigma : (R_0,...,R_n, \varepsilon_k) \rightarrow N_C \geq \frac{\varphi}{\prod_{k=1}^{a_i}(1-\varepsilon_k)} \quad (\text{eq. 6})$$

A synthesis pathway $\sigma$ is a sequence of transformations leading from an initial set of reagents $R_0$ through intermediate sets $R_0,...,R_n$ to the final product $c$. The chemputer is said to be universal if, for any target compound $c$ in the set of desired compounds $C$, there exists a sequence of transformations $\sigma$ that leads from an initial set of reagents $R_0$ to $c$.

**7. The Universal Chemputation Principle (UCP)**

Every stable, isolable molecule $c \in C$ that satisfies the abundance condition is realizable by a finite chemputation program executed on universally reconfigurable hardware where

$$\forall c \in C, \exists\ R_0 \subseteq R, P, K \text{ such that } \sigma(R_0,...,R_n) = c \text{ and } N_c \geq N_{min}(c) \quad (eq.\ 7)$$

$\sigma$ is the synthesis pathway constructed accessed using transformation function in the CSTM $\delta$

$N_c$ is the number of defect-free copies of molecule $c$ produced,

$N_{min}(c)$ is the minimum number of such copies required for the molecule to be confirmed real by detection.

This asserts that for every target compound $c$ in $C$, there exists a set of initial reagents $R_0 \subseteq R$, a set of process conditions $P$ such that a synthesis pathway $\sigma$ exists, leading from $R_0$ to $c$. This requires a dynamic error detection and correction system where

$$DEC : c' \rightarrow \text{Corrected State } c \text{ corrected} \quad (eq.\ 8)$$

Dynamic error correction (DEC) is applied at each step in the synthesis process. For each transformation, if the outcome $c'$ deviates from the expected intermediate or final product $c_n$, the error detection function DEC flags the deviation. The error correction function DEC is then applied to revert to a prior valid state or adjust the process dynamically to ensure that the synthesis remains accurate. If the compound c, is novel, then the error correction will be used to maximize the yield of the hitherto unknown compound and output a new set of rules.

The Chempiling Function ($\chi$)

$$\chi : \sigma \rightarrow G(H) \quad (eq.\ 9)$$

$\chi$ maps the synthesis pathway $\sigma$ into a hardware configuration $G(H)$ that can execute the synthesis process. For simple compounds (e.g., elements or basic molecules), the chemputer can directly synthesize them from their constituent elements or simpler precursors. If an error occurs during the synthesis of these simple compounds, it is detected and corrected dynamically before proceeding. By assuming the chemputer can synthesize all compounds of complexity $k$ (i.e., requiring $k$ steps), with dynamic error correction applied at each step. For a compound of complexity $k + 1$, there exists a precursor compound requiring $k$ steps and a transformation function $\delta$ that can transform this precursor into the target compound under appropriate $P$, and $K$ in the presence of the reagents $R$, see Fig. 5. Dynamic error correction can be used to optimise the output $\delta$ of the transformation function as it maps onto the synthesis pathway and chempiles onto the hardware. This ensures that each intermediate step is accurate. Therefore, by extension, the chemputer can synthesize all compounds up to any finite complexity, as long as there is enough compound synthesised to allow analytical detection of the molecule present in $c$.

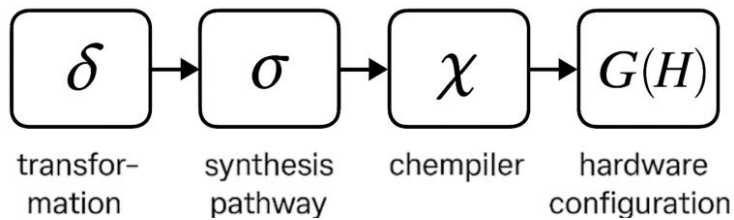

**Figure 5.** Outline of the process of chemputation from the CSTM to the synthesis pathway which then leads to chempilation on the available hardware.

The concept of dynamic error correction (DEC) represents a cornerstone of chemputation, critical to ensuring the robustness, reliability, and accuracy of fully automated chemical synthesis.[46] The foundation of DEC will lie in its ability to integrate continuous real-time analytical feedback throughout every stage of the chemical reaction, automatically identifying deviations from expected outcomes and implementing corrective measures to maintain the integrity and desired selectivity of the synthetic process. At its core, DEC will operate through a closed-loop feedback mechanism involving continuous data collection, analysis, and adaptive intervention. This cycle begins with the initial calibration of the chemputer's embedded sensors, such as spectroscopic (IR, Raman, UV-Vis, NMR), chromatographic (LC, GC, HPLC), mass spectrometric, and various physical (temperature, pressure, viscosity, conductivity) sensors. Once calibrated, these sensors will continuously acquire real-time data, can be logged and compared against the pre-defined reaction profiles embedded within the chemputer's internal reference database. As the reaction proceeds, the system will be designed such that embedded software algorithms actively monitor this sensor data, detecting deviations from expected trajectories through threshold-based or machine-learning-driven anomaly detection methods. Upon identifying a deviation, the system will classify the errors according to severity: minor deviations, indicating slight reductions in yield or selectivity; intermediate deviations, reflecting incomplete reactions or minor impurity formation; and major deviations, such as unexpected side reactions or critical intermediate failures.

Each class of error will prompt the chemputer to autonomously execute adaptive corrections tailored specifically to the severity and nature of the observed deviation. For minor deviations, subtle adjustments may be made to reaction parameters, such as incremental temperature changes or adjustments in stirring rates. Intermediate deviations will have to trigger more substantial corrections, such as additional reagent or catalyst doses, extended reaction durations, or moderate temperature adjustments. Major deviations will require more extensive interventions, such as reverting the reaction mixture to a stable precursor state and recalculating alternative reaction conditions before restarting the process. Sensor integration will be central to the effectiveness of DEC. For example spectroscopic sensors, including infrared (IR), Raman, and ultraviolet-visible (UV-Vis), could give detailed, real-time molecular information, confirming the formation of desired intermediates or products and quickly highlighting anomalies. Inline chromatographic and mass spectrometric analyses will offer precise compositional insights, rapidly verifying purity and yield at critical reaction stages. Physical sensors—such as temperature probes, pressure sensors, and pH meters—will continuously ensure that reaction environments remain within the defined optimal conditions.

It is envisaged that DEC will be further enhanced by the integration of machine-learning algorithms trained on extensive historical reaction data. Machine learning allows the chemputer to anticipate and predict potential deviations based on patterns from previous experiments, enabling proactive rather than merely reactive corrections. Moreover, these algorithms will facilitate adaptive parameter tuning, automatically refining reaction conditions over time to progressively improve yields and purity. Additionally, when significant deviations occur, machine learning algorithms may intelligently recalibrate or even redesign reaction pathways in real-time, exploring alternative synthesis routes dynamically. Thus, the careful management of error propagation will be particularly critical in multi-step syntheses. DEC effectively mitigates cumulative errors by embedding validation checkpoints after each step, systematically ensuring the quality and completeness of intermediates before progressing. Recursive correction loops further address error propagation by revisiting and correcting previous steps if downstream deviations are detected. This continual optimization of reaction parameters at each stage greatly enhances overall synthetic reliability, dramatically reducing cumulative error effects.

When it comes to exploring synthesis complexity, the assembly index $a_i$ captures the minimum causal information required to assemble a molecule. It is intrinsic to the molecule's molecular structure, independent of any synthetic route. Conceptually, $a_i$ measures how "hard" a molecule is to build in the best possible scenario: allowing for recursive reuse of parts, not just sequential chemical steps. The theoretical bounds relate the number of bonds $B$ in a molecular graph $G = (V,E)$ to the assembly index where the minimum is $log_2 B$ and the maximum is $B - 1$. Thus, molecules with a large number of bonds typically have higher $a_i$, but highly symmetric structures can have surprisingly low assembly indices due to recursive assembly. Importantly, as the assembly index increases, the synthetic error must be controlled more tightly. Otherwise, the number of correctly assembled copies declines sharply, imposing a practical limit on the size, complexity, and detectability of molecules. This effect is illustrated in Figure 6, where increasing assembly index, coupled with modest per-step error rates, leads to rapid depletion of the perfect copy population. For example, with a 5% error rate per assembly step, after only 20 steps, fewer than 40% of the molecules are flawless.

Given the inevitability of synthesis errors, dynamic error correction (DEC) is a foundational principle of chemputation. This is because DEC integrates real-time monitoring and adaptive interventions at each synthesis step. Sensors embedded within the chemputer (e.g., spectroscopy, chromatography, temperature and pH probes) continuously assess reaction progress. Deviations from expected outcomes are detected early and corrected — by adjusting reaction parameters, extending reaction time, or reverting to stable intermediates. Minor deviations may trigger fine-tuning (e.g., temperature adjustments), while major deviations may require reverting to a prior synthetic state. Machine learning algorithms further enhance DEC by predicting likely failure points based on accumulated reaction data. This closed-loop correction minimizes cumulative errors, allowing the successful synthesis of molecules with high assembly indices that would otherwise be inaccessible. It is important that we distinguish assembly index from synthetic steps.

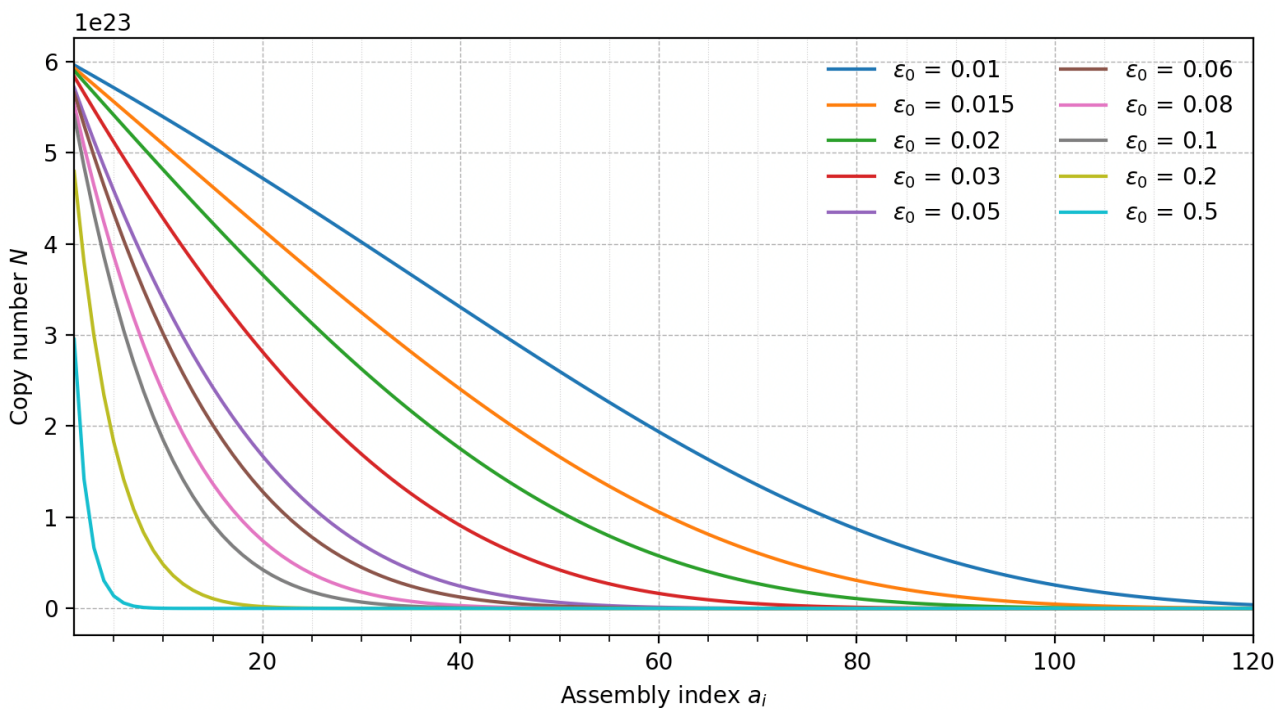

**Figure 6.** The mean number of flawless copies $N$ remaining after each assembly step ($a_i$=1 − 120) is plotted for ten baseline per-step error probabilities ($\varepsilon_0$=0.01, 0.015, 0.02, 0.03, 0.05, 0.06, 0.08, 0.10, 0.20, 0.50; coloured lines, legend right). At $a_i$=1 each system begins with Avogadro's number of copies ($N_0$=6.022×10$^{23}$). The baseline error rate rises exponentially with assembly index ($k$=0.02), and a normally distributed systematic term $E_k$ (μ=0, σ=0.005) is added to every trajectory. For each $\varepsilon_0$ value, 5,000 Monte-Carlo trajectories were generated, and the solid lines show the arithmetic mean across simulations.

A key conceptual challenge is understanding that the assembly index $a_i$ is not equivalent to the number of synthetic steps in a reaction sequence. The assembly index captures the minimal number of non-redundant construction operations required to specify a molecular graph, allowing recursive reuse of substructures. In contrast, traditional synthesis plans enumerate every experimental transformation, including redundant or linear steps, and are deeply dependent on available reagents, human planning, and practical constraints.

A lab synthesis might be "short" because it uses complex building blocks, while the true assembly index — if those blocks must themselves be constructed — is much higher. This difference is not a flaw but a feature: it reveals that assembly index is the only objective, intrinsic measure of molecular complexity. It is independent of lab strategy, human intuition, or access to reagents. In this way, assembly theory serves as an intrinsic molecular complexity, providing a universal lower bound on the information and causation needed to construct any molecule. Synthetic step count, by contrast, is contingent and observer-dependent.

A useful analogy comes from computer science. The assembly index is like the minimal program that generates an output — it's compact, elegant, and often recursive. A lab synthesis, on the other hand, is like the execution trace of that program on real hardware: long, verbose, and constrained by memory, I/O, or user habits. We wouldn't confuse a full execution trace for the "complexity" of an algorithm — similarly, we should not confuse step count with true molecular complexity. While synthetic step count remains a valuable practical metric, it cannot serve as an absolute basis for defining molecular complexity. Only the assembly index reflects the irreducible causal structure required to construct a molecule. This makes it foundational for universal chemputation, where the synthesis must be encoded, predicted, and executed independently of human bias or experimental convenience.

## 9. Examples

To explore examples of real-world synthesis done using our Chemical Description Language, χDL, we took 117 different synthesis routes we have run on our automated synthesis platforms to analyse here in terms of our CSTM abstraction[5,7,8]. This is because, in addition to directly repeating the validated procedures, this work explored the substrate scope for each χDL and showed that it can be gradually expanded by changing the substrates and adjusting key parameters – such as temperature or time - of the reaction while keeping the rest of the process unchanged. These reactions were selected based on popularity, and the resulting set of validated χDLs covers a substantial range of common reactions and constitutes an entry point to the automation the entire organic synthesis 'toolbox'. Also, all the procedures cover highly diverse chemistry showing that one unified hardware and software of the Chemputer can indeed work, see Figure 7.

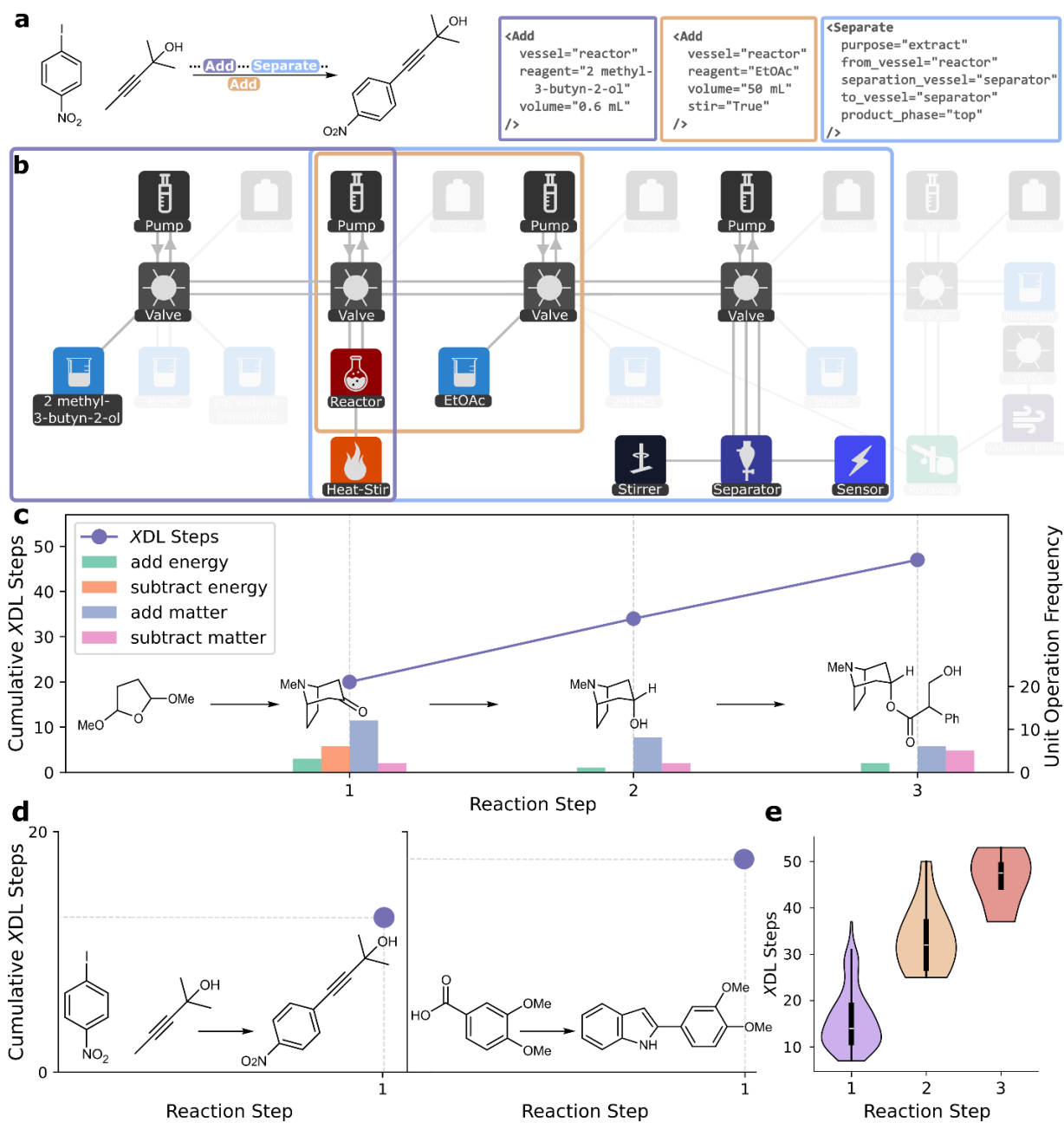

**Figure 7:** Graphs showing the scaling behaviour of χDL steps, according to our CSTM abstraction, as a function of reaction steps. **a)** Reaction schema of 2-Methyl-4-(4-nitrophenyl)but-3-yn-2-ol and three exemplary χDL steps from the respective synthesis. **b)** Abstract chemputer graph showing the actively addressed component for three χDL steps, as indicated in a). **c)** Scaling behaviour of the cumulative number of χDL steps for the three-step synthesis of Atropine. The distribution of unit operations per reaction step is shown on the secondary y-axis. **d)** Number of χDL steps for the two single-step reactions of 2-Methyl-4-(4-nitrophenyl)but-3-yn-2-ol, and 2-(3,4-dimethoxyphenyl)-1H-indole. **e)** Distribution of the cumulative number of χDL steps for single-step, two-step, and three-step reactions, respectively.

Finally, the exploration of these previously done reactions on well-established robotic systems, highlights how this approach can be used to unify and also standardise the use of diverse robotic architectures for chemical automation. While synthesis for all molecules can be expressed in using a universal set of hardware components (see above), each synthesis(-step) may employs a unique combination of resources. This is exemplified with on three χDL steps in the synthesis of 2-Methyl-4-(4-nitrophenyl)but-3-yn-2-ol (Figure 7a/b). Additionally, as illustrated in Figure 7c using the example of atropine synthesis, the total number of required χDL steps increases approximately linearly with the number of reaction steps (20, 34, 47). Each synthesis step can be fully captured using our CSTM abstraction. This is further demonstrated in Figure 7d with two single-step syntheses—2-(3,4-dimethoxyphenyl)-1H-indole and 2-methyl-4-(4-nitrophenyl)but-3-yn-2-ol—which require 18 and 13 χDL/CSTM abstraction steps, respectively. An analysis of 117 physically executed synthesis routes, covering a broad and representative range of chemical transformations, confirms the practical utility and generalizability of our abstraction for driving automated synthesis platforms and self-driving laboratories (Figure 7e). These findings also highlight that even complex, multi-step syntheses can be expressed through a linearly scaling number of unit operations as defined in section 3 (see Table S1 for classification of χDL steps).

## 10. Practical Limitations

Implementing the concept of chemputation in practice presents a series of significant challenges that extend beyond this robust theoretical framework. One of the foremost challenges lies in the complexity and scalability of the chemputer's hardware. The concept of universally configurable hardware, which is central to the chemputer's ability to synthesize any chemical compound, demands a highly versatile and flexible system with a range of different modules for operations like filtration, extraction and so on. Designing hardware that can seamlessly switch between different configurations for a wide variety of chemical processes is an intricate task. Each module within the system must handle diverse reaction types, process conditions, and scales of operation while maintaining precision and reliability. Moreover, there is an inherent tension between the need for miniaturization, which allows for precision, and the requirement for scalability to manage larger volumes or more complex reactions. Achieving both in a single system, particularly one that remains flexible and configurable, is a significant engineering challenge. Furthermore, the integration of this hardware with the software responsible for the chempiling function—mapping synthesis pathways to specific hardware configurations—adds another layer of complexity. This software must dynamically adjust the hardware setup in real-time, requiring a level of synchronization and control that is difficult to achieve.

One critical new point introduced in this work is the expansion of the very definition of a molecule by explicitly linking its synthetic accessibility[50–52] to the intrinsic complexity of its molecular graph, quantified through the concept of the assembly index.[40,42,43] Traditionally, chemists have viewed synthetic complexity through heuristic measures, such as the number of reaction steps or yield, without a fundamental underlying theoretical framework. While mapping synthetic complexity onto an abstract concept like the assembly index might initially seem non-intuitive, doing so provides a rigorous theoretical grounding. Specifically, the assembly index measures the minimal number of causal steps or constraints required to produce a molecule, giving a precise quantification of molecular complexity directly related to synthesis. By establishing a clear link between molecular complexity and synthesis, this approach effectively sets an absolute lower bound on the number of parameters needed to orchestrate and control the synthetic process. Thus, rather than relying purely on empirical knowledge or reaction heuristics, chemists can now use the assembly index to systematically define, compare, and predict the synthetic accessibility and complexity of molecules. This theoretical advance not only facilitates better predictive models for chemical synthesis but also deepens the fundamental understanding of molecular construction, bridging the gap between theoretical complexity measures and practical chemical synthesis.

Another critical challenge is the implementation of dynamic error correction within the chemputer, which is essential for ensuring the accuracy and reliability of chemical syntheses. The system must be capable of real-time monitoring and adjustment, continuously tracking the progress of each reaction, detecting any deviations from the expected pathway, and applying corrective measures immediately. Recently we have shown that systems with in-line sensors and dynamic χDL can be used for both error correction and optimisation.[46] To achieve this demanded advanced sensing technologies and real-time data processing capabilities that can operate effectively across a broad range of reaction conditions. However, in multi-step syntheses, errors can propagate through the system, compounding and becoming more difficult to correct as the process continues. Developing mechanisms that can effectively manage and contain such errors, ensuring the robustness and redundancy of the system, is crucial. Achieving this balance between robustness, cost, space, and energy efficiency poses a significant challenge. The theoretical framework also assumes a comprehensive understanding of the chemical space and the ability to encode all possible reactions into the chemputer. However, the reality of chemical synthesis is more complex. Our current knowledge of chemical reactions is not exhaustive, particularly in the fields of complex organic and biological chemistry, where many reactions remain poorly understood or unpredictable. This limitation restricts the chemputer's ability to reliably handle all potential syntheses. Moreover, as complex molecules are synthesized, emergent properties may arise that are not predicted by existing models, leading to unexpected reactions or products. The chemputer must be designed to manage and correct such deviations, even in the face of novel or poorly understood chemistry. Developing algorithms and hardware that can adapt to new chemical data in real-time is a significant hurdle that must be overcome.

**Conclusions**

Since the chemputer, as seen as the CSTM can implement any transformation function $\delta$ and can control all relevant process conditions, it can instantiate any chemical synthesis process. The inclusion of dynamic error detection and correction at each step ensures the reliability and accuracy of the synthesis. Additionally, the use of universally configurable hardware and the chempiling function allows the chemputer to dynamically adapt its configuration for various synthesis pathways. Thus, the chemputer is universal for chemical synthesis, capable of generating any compound $c \in C$ given the appropriate initial conditions, transformations, and error correction mechanisms. The formalization above establishes the concept of a chemputer as a universal chemical synthesis machine. The CSTM transformation function $\delta$, synthesis pathways $\sigma$, stability conditions $S$, dynamic error correction (DEC), chempiling function $\chi$, and configurable hardware $H$ together define a universal model capable of synthesizing any target compound within the chemical space defined by $R, P$. The work presented here establishes the chemputer as a universal chemical synthesis machine, demonstrating its capability to synthesize any target compound within a defined chemical space by the process of chemputation. By formalizing the key components, such as the transformation function $\delta$, synthesis pathways $\sigma$, stability conditions $S$, dynamic error correction DEC, and the chempiling function $\chi$, we have constructed a robust theoretical framework that underpins this universality. The integration of universally configurable hardware further enhances the chemputer's adaptability, allowing it to dynamically reconfigure and execute a wide array of chemical processes with precision. This is universal considering finite constraints on the reaction hardware, reagents, reaction steps, reaction time, and the ability to produce the target molecule in a detectable amount.

**Supplementary Information**

The CSTM is described in detail in the supplementary. A Mathematica notebook is available, data and python scripts for figures 4 and 5, and a supplementary video (SV1) showing how the CSTM works on a 1D tape.

https://www.dropbox.com/scl/fo/hbntfkivkhxdm4p1axzek/AN1feY7IPn2N5sYcmivF1XU?rlkey=stbnxq17z6n14bv5z1dd3tvd9&dl=0

## Acknowledgements

We would like to thank David Deutsch, Muffy Calder, Sara Walker, S. Hessam Mehr, Keith Patarroyo, Emma Clarke, Dario Caramelli, and Edward Lee for comments and feedback. We acknowledge financial support from the John Templeton Foundation (grants 61184 and 62231), Sloan Foundation, Schmidt Futures, NIH, Google, and EPSRC (grant nos. EP/L023652/1, EP/R01308X/1, EP/S019472/1, and EP/P00153X/1 and ERC (project 670467 SMART-POM).

## Author Contributions

LC conceived the idea, did the proofs, the assembly analysis and formalised the CSTM. AS helped validate the formalism. SP expanded the experimental analysis with advice from LC and AS. The manuscript and SI was written by LC with suggestions from AH and SP.

## Competing interest statement

The authors declare that they have no competing financial interests.